\makeatletter
\declare@file@substitution{revtex4-1.cls}{revtex4-2.cls}
\makeatother

\documentclass[twocolumn,preprint]{aastex631}
\usepackage[utf8x]{inputenc}
\usepackage{amsmath,amsfonts,amssymb,mathrsfs,bm,enumerate,graphicx}
\usepackage[T1]{fontenc}
\usepackage[all]{hypcap}
\hypersetup{
    colorlinks=true,
    linkcolor=magenta,
    filecolor=magenta,      
    urlcolor=blue,
}

\usepackage{enumitem} 

\usepackage{ulem}
\usepackage{threeparttable}
\usepackage{ae,aecompl}
\usepackage{wasysym}
\usepackage{booktabs}

\def\red#1 {\textcolor{red}{#1}\ }   
\def\blue#1 {\textcolor{blue}{#1}\ }   
\def\purple#1 {\textcolor{purple}{#1}\ }   

\def\red#1 {\textcolor{red}{#1}\ }   
\def\blue#1 {\textcolor{blue}{#1}\ }   

\usepackage{ulem}
\usepackage{threeparttable}
\usepackage{ae,aecompl}
\usepackage{wasysym}
\usepackage{xcolor}
\definecolor{orange}{RGB}{255,127,0}
\definecolor{pinegreen}{HTML}{009B55}

\makeatletter
\newcommand{\shorteq}{\mathrel{\mkern0.2mu\mathpalette\shorteq@\relax\mkern0.2mu}}
\newcommand{\shorteq@}[2]{\scalebox{0.5}[1]{$\m@th#1=$}}
\DeclareMathSymbol{\shortminus}{\mathbin}{AMSa}{"39}

\shorttitle{Planet--Planet Secular Migration Predicts a Stellar Obliquity--Period Anti-Correlation}
\shortauthors{Bhaskar, Petrovich \& Mu\~noz}

\begin{document}



\title{Planet--Planet Secular Migration Predicts a Stellar Obliquity--Period Anti-Correlation}

\author[0000-0002-5181-0463]{Hareesh Gautham Bhaskar}
\affiliation{Indiana University, Bloomington, Indiana, USA}
\affiliation{Technion-Israel Institute of Technology, Haifa, Israel}
\author[0000-0003-0412-9314]{Cristobal Petrovich}
\affiliation{Indiana University, Bloomington, Indiana, USA}
\author[0000-0003-2186-234X]{Diego J.\ Mu\~noz}
\affiliation{Department of Astronomy and Planetary Science, Northern Arizona University, Flagstaff, AZ 86011, USA}

\begin{abstract}

Stellar obliquities provide a fossil record of hot Jupiter (HJ) migration. An emerging observational trend in single-star systems is that strongly misaligned HJs are largely confined to short orbital periods, while longer-period HJs are preferentially aligned.
This pattern cannot be explained by tidal dissipation in the star and may instead preserve clues to the migration pathway. We show that secular high-eccentricity migration driven by a distant planetary companion naturally produces such an obliquity--period correlation. In our simulations, the shortest-period HJs tend to be produced by the von Zeipel--Lidov--Kozai mechanism driven by highly inclined companions, which results in a broad range of final stellar obliquities. The longest-period HJs, on the other hand, are produced over longer timescales by coplanar high-eccentricity migration, which preserves low obliquities. The transition between these two limits is not abrupt, with intermediate-period HJs displaying a moderate range of obliquities.
According to this interpretation, we predict that the shortest-period HJs should have distant planetary companions with broadly distributed mutual inclinations, whereas the companions of longer-period HJs should reside in nearly coplanar orbits. Upcoming Gaia astrometric constraints will provide a key test of this picture.

\end{abstract}


\section{Introduction}

Soon after the discovery of the first hot Jupiter (HJ)\textemdash51 Peg~b\textemdash tidal dissipation in eccentric, low-periastron giant planets was recognized as an crucial driver of orbital evolution \citep[e.g.,][]{Rasio:1996a,Weidenschilling:1996}. The shrinking of the planet's semi-major axis by up to 2 orders of magnitude could thus be made very effective,
provided that an eccentricity excitation mechanism were present. This process
is now broadly referred to as ``high-eccentricity migration'' \citep[HEM; e.g.,][]{Socrates:2012}, and it encompasses a number of pathways that lead to eccentricity growth.

The diverse channels or ``flavors'' of HEM include planet-planet scattering \citep{Rasio:1996b,nagasawaFormationHotPlanets2008,beaugeMultipleplanetScatteringOrigin2012b}, von Zeipel-Lidov-Kozai (ZLK) oscillations with tidal friction \citep{wuPlanetMigrationBinary2003,fabryckyShrinkingBinaryPlanetary2007},
secular chaos \citep{wuSecularChaosProduction2011a,LithwickWu2014},
and coplanar high-eccentricity migration \citep[CHEM;][]{petrovichHotJupitersCoplanar2015}.
All these appear effective at producing HJs, which can make the identification of a preferred formation channel uncertain. Nonetheless, Monte Carlo experiments show that
the resulting distribution of orbital properties can be unique to each migration channel. Notably, all these mechanism produce a distinct
distribution of spin-orbit misalignments \citep{AlbrechtObliquity2022}, and thus the observed distribution of stellar obliquities can in principle be used to constrain the dominant HJ migration channel \citep[e.g.,][]{Morton:2011,Rice:2022}

Historically, observational bias has made the accurate determination of stellar obliquities difficult for all but the shortest-period HJs. Modern spectrographs, however, have pushed these detections toward smaller and longer-period planets, enabling spin-orbit measurements of warm jupiters \citep{Rice2022_WJs,JIR_2025_Jupiters} and even sub-Neptunes \citep{Polanski:2025}. These advances now enable the statistical study of the stellar obliquity in HJ systems as a function of orbital period, revealing a tentative trend that HJs on closer-in orbits (\( a < 0.06\,\mathrm{au} \)) display a broader range of stellar obliquities compared to those on wide orbits (\( a = 0.06-0.1\,\mathrm{au} \)).  This pattern cannot be attributed to tidal realignment, as tidal forces are stronger for close-in planets and are expected to realign their orbital planes with the stellar spin axis. The broader obliquity distribution among wider-orbit Jupiters may therefore reflect differences in their dynamical histories and migration mechanisms.

\begin{figure}
	\centering
	\includegraphics[scale=0.4]{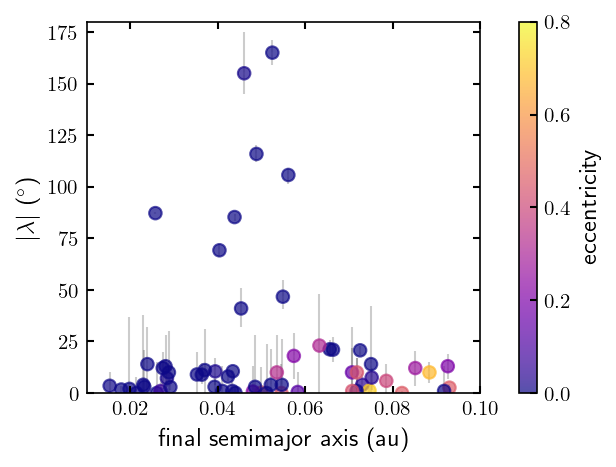}	
	\caption{The observed projected obliquity distribution of HJs (with mass $>1\,M_\mathrm{Jup}$) around single stars (with mass $>0.5\,M_\odot$). The final semi-major axis of the planets ($a_{\rm f}= a[1-e^2]$) are shown on the x-axis, and their projected obliquities are shown on the y-axis. The colors show the eccentricities of the planets. In general, planets on close-in orbits ($a_{\rm f}<0.06$ AU) have low eccentricity ($e \sim 0.1 $) and broad range of projected obliquities ($0 < \lambda < 180^\circ$).  Meanwhile, planets on wider orbits ($a_{\rm f}>0.06$ AU), tend to be more eccentric ($ e \sim 0.3 $), and have a narrow range of obliquities ($\lambda < 50^\circ$).}
	\label{fig:ensa1fpsif}
\end{figure}

We illustrate this trend in Figure~\ref{fig:ensa1fpsif}, where the projected obliquity $\lambda$ is plotted against the ``final'' semi-major axis
$a_{\rm f}\equiv a(1-e^2)$. Not only does obliquity depend strongly $a_{\rm f}$, but so does
the eccentricity of the planets, which tend to be higher at wider separations, in alignment with the expectation that tidal dissipation has not had enough time to circularize the orbits at these wider separations \citep{JIR_2025_Jupiters}.

In this work, we demonstrate that the observed correlation between stellar obliquity and the orbital properties of HJs can be naturally explained by their migration history through planet–planet secular HEM (PPSHEM). In particular, we show that the dichotomy in the obliquity distribution arises because closer-in HJs predominantly form via conventional ZLK-driven migration, whereas wider-orbit HJs are more likely produced through coplanar high eccentricity migration (CHEM), which naturally yields lower stellar obliquities. The transition between these two regimes is smooth and governed by their respective migration timescales, with CHEM  occurring over longer timescales.

In literature, PPSHEM has been generally studied under two limits: the (eccentric) ZLK limit, in which mutually inclined planets undergo HEM \citep{Naoz2011,Teyssandier2013,Lu2025}, and CHEM \citep{Li2014,petrovichSteadystatePlanetMigration2015,Xue2017_chem} which is triggered in  systems with eccentric coplanar planets. It should be noted that the underlying dynamics of both of these regimes is driven by the same secular Hamiltonian \citep[e.g.,][]{Naoz2013}. In this work we sample the entire range of initial conditions in both eccentricities and mutual inclinations which not only includes both of these limits, but also the intermediate parameter space characterized by moderate eccentricities and inclinations.

\section{Secular Dynamics Modeling}
\label{sec:results}
\subsection{Setup}
\label{sec:setup}

We investigate the secular dynamics of hierarchical two-planet systems orbiting a central, spinning star, including tidal dissipation acting on the inner planet. Similar numerical experiments have been performed in previous studies \citep{Naoz2012,petrovichHotJupitersCoplanar2015}; a detailed description of our setup is provided in \S~\ref{sec:setup_appendix}.

In brief, our model incorporates short-range forces arising from rotational and tidal deformations, as well as general relativistic precession. Tidal dissipation is implemented using an equilibrium-tide framework. The planets are widely separated, allowing the dynamics to be modeled with a double-averaged Hamiltonian expanded to octupole order \citep{naozSecularDynamicsHierarchical2013}.

The initial conditions for our fiducial simulations are as follows:
\begin{itemize}
\item The inner and outer planets are initialized with semi-major axes of 1~au and 10~au, respectively.
\item The inner planet begins with zero obliquity relative to the stellar spin axis, while the outer planet’s orbital orientation (i.e., mutual inclination) is sampled on a grid with $\Delta I_\mathrm{mut}=3^\circ$.
\item The orbital eccentricities are sampled on a uniform grid with $\Delta e=0.05$ \footnote{We use a uniform grid in eccentricity and mutual inclination, rather than sampling these parameters randomly, to facilitate reanalysis of our fiducial simulations under different weighted initial eccentricity and inclination distributions.}.
\end{itemize}
Since eccentric mutually inclined planetary systems can be unstable \citep{Hadden2018,Bhaskar2024}, systems that do not satisfy the stability criterion of \citet{Petrovich2015} are discarded. We stop our simulations either when a stop time of $10^{10}$ years is reached, the planet is tidally disrupted \footnote{Following \cite{Guillochon2011}, we assume that the inner planet is disrupted when $a_1(1-e_1)<2.7R_{\odot}$.} or when the orbit of the HJ is circularized ($e<0.1$).

Our strategy is to remain agnostic about the initial eccentricities and inclinations, without imposing a priori weights that would favor systems evolving through either ZLK or CHEM channels. To this end, we performed a large ensemble of simulations ($N_{\rm systems}=16556$), enabling us to re-sample the results under more realistic distributions in post-processing (e.g., Rayleigh distributions for inclinations and eccentricities).

\subsection{Results}
\label{sec:fidres}
\begin{figure*}
	\begin{center}
	\includegraphics[scale=1]{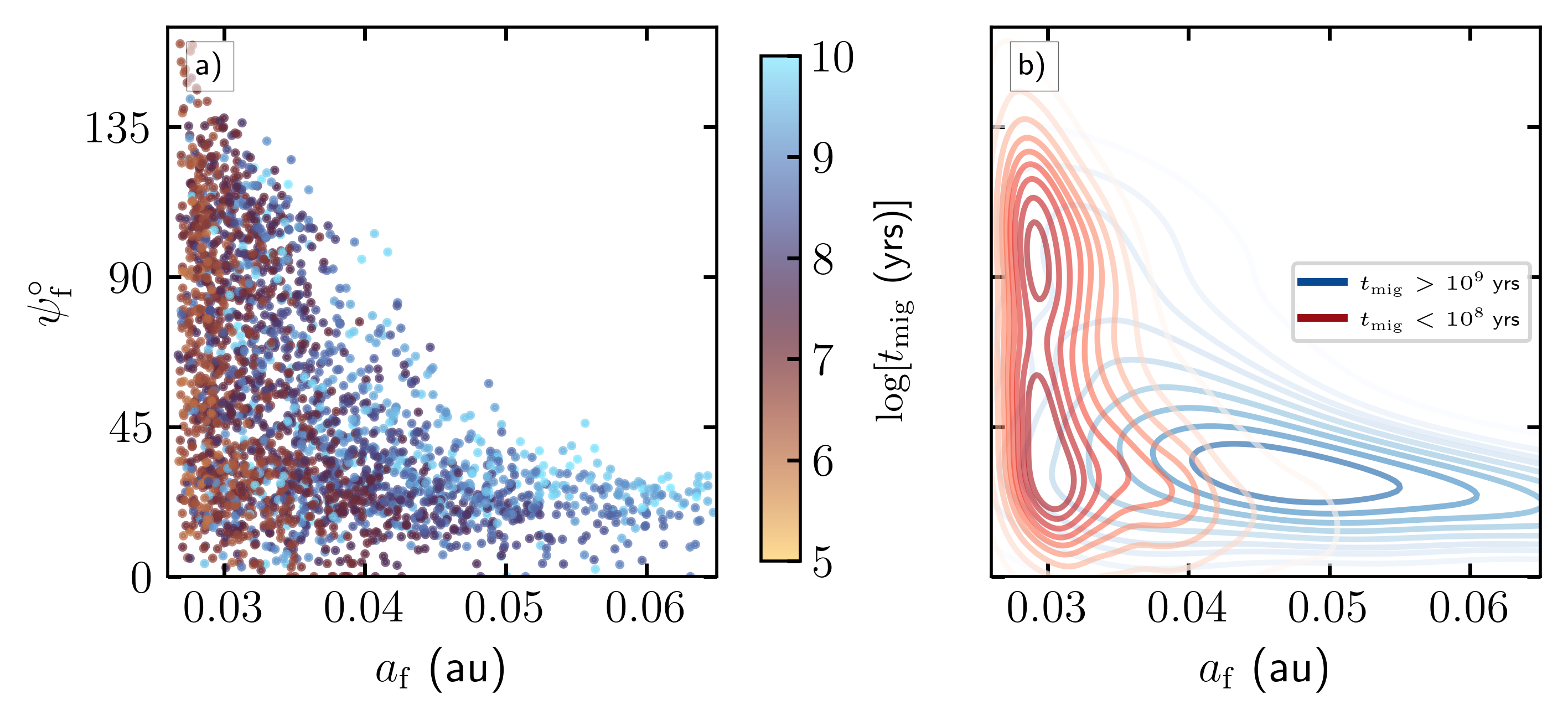}
    
	\caption{Final semi-major axes and obliquities of HJs formed in our fiducial simulations. Panel (a): the final semi-major axis of HJs are shown on the x-axis, and the final obliquities are shown on the y-axis. The color indicates the migration timescale. Close-in HJs exhibit a broad range of obliquities, whereas wider-orbit HJs tend to have low obliquities. The distribution shows an envelope that constrains the maximum obliquity of HJs at a given final semi-major axis. Panel (b): kernel density estimates of fast- and slow-migrating HJs in \(a_{\mathrm{f}}\)-\(\psi_{\mathrm{f}}\) space. HJs that take longer to migrate tend to have lower obliquities and wider orbits, whereas rapidly migrating HJs tend to end up on close-in orbits with a broad range of obliquities.}
	\label{fig:fidsimres}
	\end{center}
\end{figure*}

Our Monte Carlo simulations produce 2380 hot and warm Jupiters with $a_\mathrm{f}(1 - e_\mathrm{f}^2)<0.1$ au that would end as a HJ provided enough time to circularize. In this section, we present the results of our fiducial simulations. 

Figure~\ref{fig:fidsimres} illustrates the properties of planets that underwent HEM. In panel (a), we show the final semi-latus rectum of the planet ($a_{\rm f}$) on the x-axis and the final obliquity ($\psi_{\rm f}$) on the y-axis, with color indicating the time required for each planet to migrate to its close-in orbit. Similar to the observed trend, we can see that close-in planets tend to span a broad range of obliquities, whereas wide-orbit planets generally have low obliquities.

The importance of octupole-order terms in the secular Hamiltonian can be quantified by the parameter $\epsilon = a_1 e_2/[a_2(1-e_2^2)]$. We find that wide-orbit HJs were typically initialized with moderate values of $\epsilon$ ($\sim 0.05$) and low initial mutual inclinations ($I_{\rm mut,init} \lesssim 30^\circ$), whereas close-in HJs often originate with lower $\epsilon$ values ($\sim 0.01$) and near-polar inclinations ($I_{\rm mut,init} \sim 90^\circ$). In the intermediate regime, the dynamics become more complex, and no simple correlation emerges between the initial inclination and final obliquity.

This behavior suggests that close-in HJs primarily undergo HEM driven by quadrupole-order ZLK oscillations, whose shorter-timescale perturbations deliver planets to tight orbits. Conversely, wide-orbit HJs undergo migration dominated by octupole-order perturbations, previously identified as CHEM. The resulting obliquity distributions are consistent with previous studies: ZLK-driven HEM produces a broad range of obliquities \citep{fabryckyShrinkingBinaryPlanetary2007}, whereas coplanar HEM yields low obliquities \citep{petrovichHotJupitersCoplanar2015}. Thus, the observed correlation between obliquity and semi-major axis in HJs is well reproduced by our fiducial simulations.

We also find an obliquity envelope that constrains the maximum obliquity HJs can attain for a given final semi-major axis. A similar trend appears in Figure 10 of \cite{fabryckyShrinkingBinaryPlanetary2007}, where the authors considered only quadrupole-order ZLK-driven HEM. In Appendix \ref{app:envder}, we derive an analytical expression for this envelope in the quadrupole limit. Beyond this limit, which is valid only for close-in HJs with near-polar companions, the envelope extends continuously to lower inclinations.

The color scale indicates the migration time: close-in planets tend to migrate faster, whereas wider-orbit planets migrate more slowly. This trend is shown explicitly in panel (b), which presents kernel density estimates for rapidly migrating planets ($t_\mathrm{mig}<10^8$ yr) and slowly migrating planets ($t_\mathrm{mig}>10^9$ yr). This behavior is consistent with the interpretation that close-in planets migrate through quadrupole-order ZLK oscillations, while wider-orbit planets migrate through longer-timescale octupole-order CHEM.

\subsection{Dependence on initial conditions}
\label{sec:ICs}


Our focus on PPSHEM is broadly consistent with previous planet–planet scattering studies, which have shown that HJs formed following dynamical instabilities predominantly migrate through high-eccentricity pathways, often involving a combination of secular processes rather than a single migration mechanism \citep[e.g.,][]{Garzon2022}.

This diversity is naturally captured by our agnostic initial conditions, which sample a broad range of eccentricities and inclinations. Our choice of initial semi‑major axes is also consistent with planet–planet scattering outcomes: after the initial instability, the remaining planets typically exhibit a semi‑major axis ratio of $\sim\,10$, with the inner planet residing at a few au distance \citep[e.g.,][]{nagasawaFormationHotPlanets2008}.


In our fiducial simulations, we sample the initial eccentricities and mutual inclinations on a uniform grid. The correlation between the final obliquity and semi-major axis shown in Figure \ref{fig:fidsimres} arises from the underlying secular dynamics, but it cannot be directly compared to observations due to uncertainties in the initial eccentricity and inclination distributions. Hence, we reanalyze our fiducial simulations by performing a weighted kernel density estimation in the $a_\mathrm{f} - \psi_\mathrm{f}$ space.

\begin{figure*}
\centering
	\includegraphics[scale=1.0]{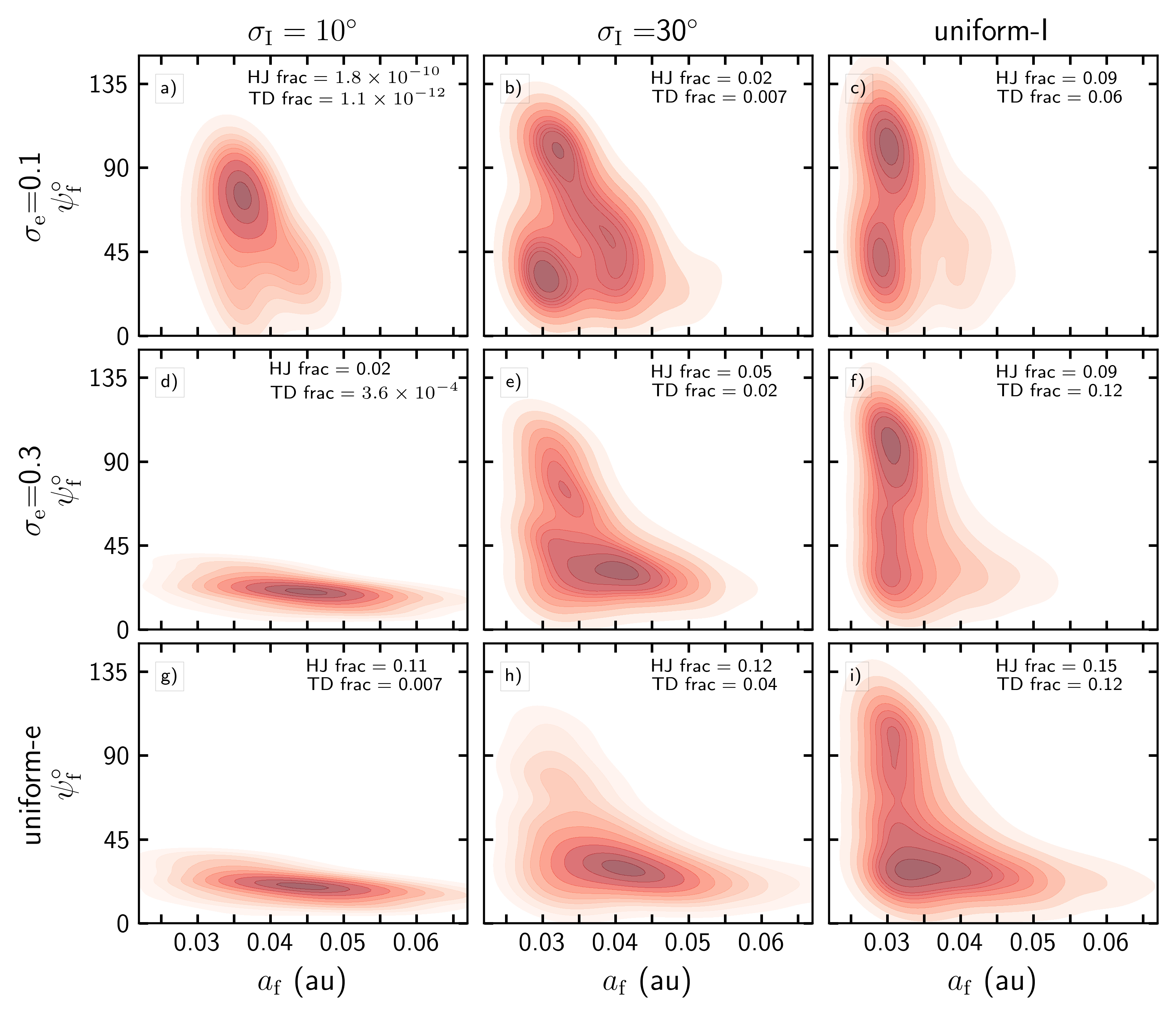}
	\caption{ Weighted distributions of the final semi-major axes and obliquities of HJs formed in our simulations. In our fiducial simulations, the initial eccentricities and mutual inclinations of the planets are sampled on a uniform grid. In this figure, we assign weights to each HJ to reflect different assumed underlying distributions of initial eccentricities and mutual inclinations. Specifically, we compute the weights using Rayleigh distributions with varying scale parameters, \(\sigma\), as well as uniform distributions. The assumed distribution of initial eccentricities varies between rows, while the assumed distribution of initial mutual inclinations varies between columns. When the initial mutual inclination distribution is strongly weighted toward low values, only CHEM is allowed. In contrast, when the initial mutual inclination distribution is broad and the initial eccentricity distribution is weighted toward low values, the ZLK mechanism dominates. Sufficiently high initial eccentricities and mutual inclinations ($\sigma_{\rm e}>0.3$ and $\sigma_{\rm I}>30^\circ$) are required to recover the semi-major-axis--obliquity correlation found in our fiducial simulations.}
    \label{fig:afpsifweights}
\end{figure*}

In Figure~\ref{fig:afpsifweights}, we present the weighted probability density distributions of $a_\mathrm{f}$  and $\psi_\mathrm{f}$  obtained from our fiducial simulations. These distributions are computed by assigning weights to each simulation in our ensemble. The weights reflect the assumed probability distributions of the planets’ initial eccentricities and mutual inclinations. We adopt Rayleigh distributions with varying scale parameters, as well as uniform distributions. Across columns, we vary the initial mutual-inclination distribution, while across rows we vary the initial eccentricity distribution. In the top-right corner of each panel we report the fraction of systems that form HJs and those that undergo tidal disruption. This analysis is designed to assess how sensitive our simulation outcomes are to the assumed initial conditions.

In panel (a) the initial mutual inclinations between the planets and eccentricities are too low for either CHEM or ZLK-driven oscillations to operate efficiently. Consequently the weighted HJ formation fraction is too low. The planets which do migrate in this regime tend to have low initial eccentricities ($e_\mathrm{1,init}\sim0.1$) with companions initialized with moderate eccentricities ($e_\mathrm{2,init}\sim0.3-0.5$), and mutual inclinations ($\sim55^\circ$). This corresponds to neither ZLK-driven HEM, nor CHEM, but an intermediate mechanism driven by octupole order perturbations. We find that planet which comprise the envelope seen in Figure \ref{fig:fidsimres} migrate through this intermediate mechanism.

If the initial eccentricities of the planets are higher (panels d and g), CHEM becomes important, mainly producing low obliquity wide orbit HJs. HJ fraction and tidal-disruption fraction  increases as the initial eccentricity distribution shifts to larger values. Shifting the inclination distribution to larger values activates the ZLK mechanism as well (panels b and c). This populates the close-in high obliquity parameter space. In panel (i) we recover the fiducial distribution seen in Figure \ref{fig:fidsimres}.  Overall, we need $\sigma_\mathrm{e} \gtrsim 0.3$ and $\sigma_\mathrm{I} \gtrsim 30^\circ$ to reproduce the observed correlation between the final semi-major axis and obliquities of HJs. 

\begin{figure*}
\centering
    	\includegraphics[scale=1.2]{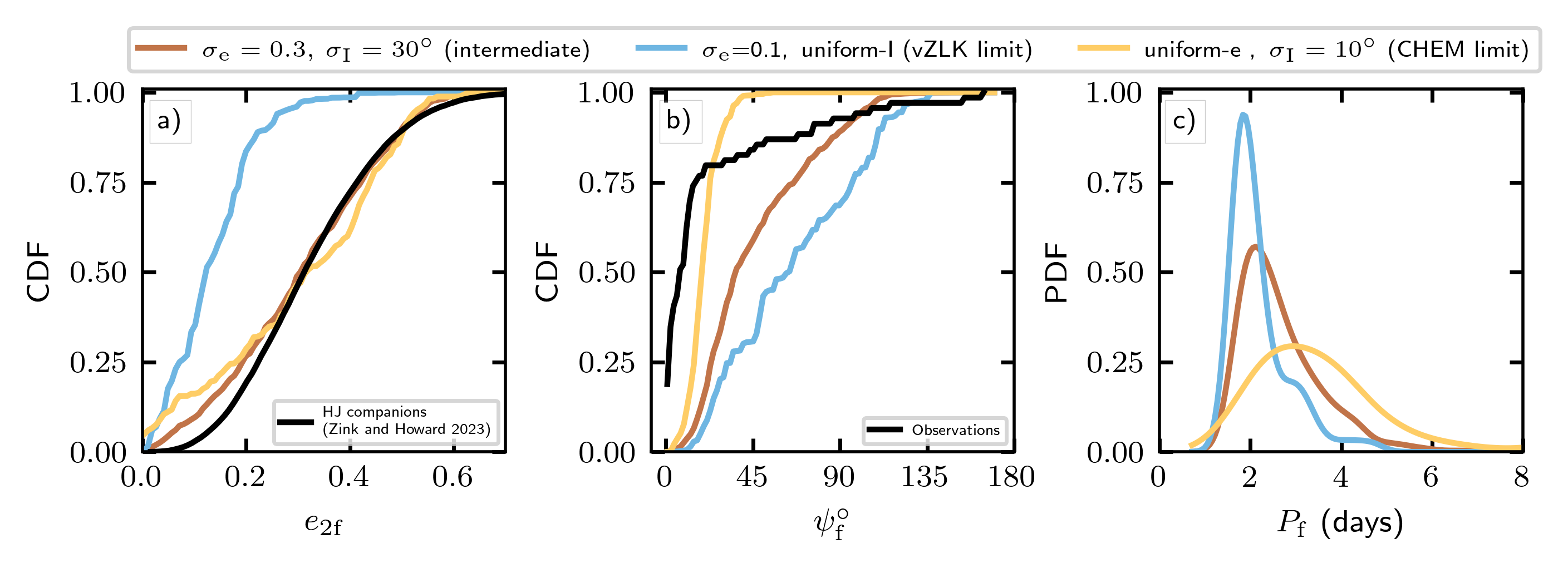}
	\caption{Marginalized distributions of companion eccentricity (panel a), HJ obliquity (panel b), and final orbital period (panel c) from our simulations. The colored curves correspond to different assumed underlying distributions of the planets' initial eccentricities and mutual inclinations. We focus on three regimes: the ZLK limit, the CHEM limit, and an intermediate case. In panel (a), the black curve shows the observed eccentricity distribution of HJ companions from \citet{zinkHotJupitersHave2023a}. In panel (b), the black curve shows the obliquity distribution from Figure~\ref{fig:ensa1fpsif}. The observed companion eccentricity distribution is consistent with both the intermediate case and the CHEM limit. The observed obliquity distribution contains more aligned systems than our simulations. The final-period distribution depends on the relative contributions of the different migration channels: the ZLK limit peaks at the shortest periods, the CHEM limit peaks at the longest periods, and the intermediate case peaks between these two limits.}
    \label{fig:e2cdf}
\end{figure*}

\subsubsection{Marginalized distributions: companion eccentricities, obliquities and periods}
We now analyze the broad properties of HJs and their companions. We focus on three representative choices for the initial eccentricity and mutual inclination distributions. First, we include the weighted distribution with uniform initial inclinations and $\sigma_e=0.1$, which we refer to as the ZLK limit. Second, we consider the weighted distribution with low initial inclinations, $\sigma_\mathrm{I}=10^\circ$ and uniform initial eccentricities, which we refer to as the CHEM limit. Finally, we include an intermediate case with $\sigma_\mathrm{e}=0.3$ and $\sigma_\mathrm{I}=30^\circ$, corresponding to panel (e) of Figure \ref{fig:afpsifweights}.

In panel (a) of Figure \ref{fig:e2cdf}, we show the final eccentricity distributions of the outer companions to HJs. The black curve shows the observed distribution from \cite{zinkHotJupitersHave2023a}, who found an average companion eccentricity of 0.34 and interpreted this as evidence that CHEM is a dominant HJ formation pathway. Our results are consistent with this interpretation: the ZLK limit produces companion eccentricities below the observed values, whereas the CHEM limit yields higher eccentricities that are broadly consistent with the observations. The intermediate case closely follows the CHEM-limit distribution and is also consistent with the observed distribution.

In panel (b), we compare the final obliquity distributions for the same three initial conditions. The observed distribution from Figure \ref{fig:ensa1fpsif} is shown in black. In the CHEM limit, most HJs have low obliquities, whereas in the ZLK limit, HJs tend to reach higher obliquities. The intermediate case lies between these two limits. However, the observed obliquity distribution contains an overabundance of low-obliquity systems relative to all three simulated distributions. This discrepancy may be due to tidal realignment of close-in, ZLK-driven HJs, which initially tend to have higher obliquities. Because tidal realignment predominantly affects close-in HJs, it can reduce their obliquities and thereby increase the low-obliquity population \citep[e.g.,][]{Zanazzi2024}. In addition, other formation channels, such as in-situ formation or migration, are also expected to contribute predominantly to the low-obliquity population, but these channels are not included in this study.

In panel (c), we show the period distributions of HJs produced by the three initial-condition distributions. All three cases yield peaked period distributions, but the location of the peak depends on the dominant migration channel. The ZLK limit peaks at shorter orbital periods with a median orbital period of 1.9 days, whereas the CHEM limit peaks at wider orbits with a median orbital period of 3.4 days. The intermediate case peaks between these two limits with a median orbital period of 2.4 days. The smaller secondary peak in the ZLK limit is due to a contribution from CHEM. The observed HJ period distribution is also peaked, with a characteristic peak near 4 days \citep[e.g.,][]{Yee2023}. Nevertheless, a direct comparison with the observed period distribution is not straightforward because the planetary tidal quality factor, which is poorly constrained, controls the final semi-major axis in our simulations (see Appendix \ref{sec:tidemodel}).

\begin{figure*}
\centering
	\includegraphics[scale=1.2]{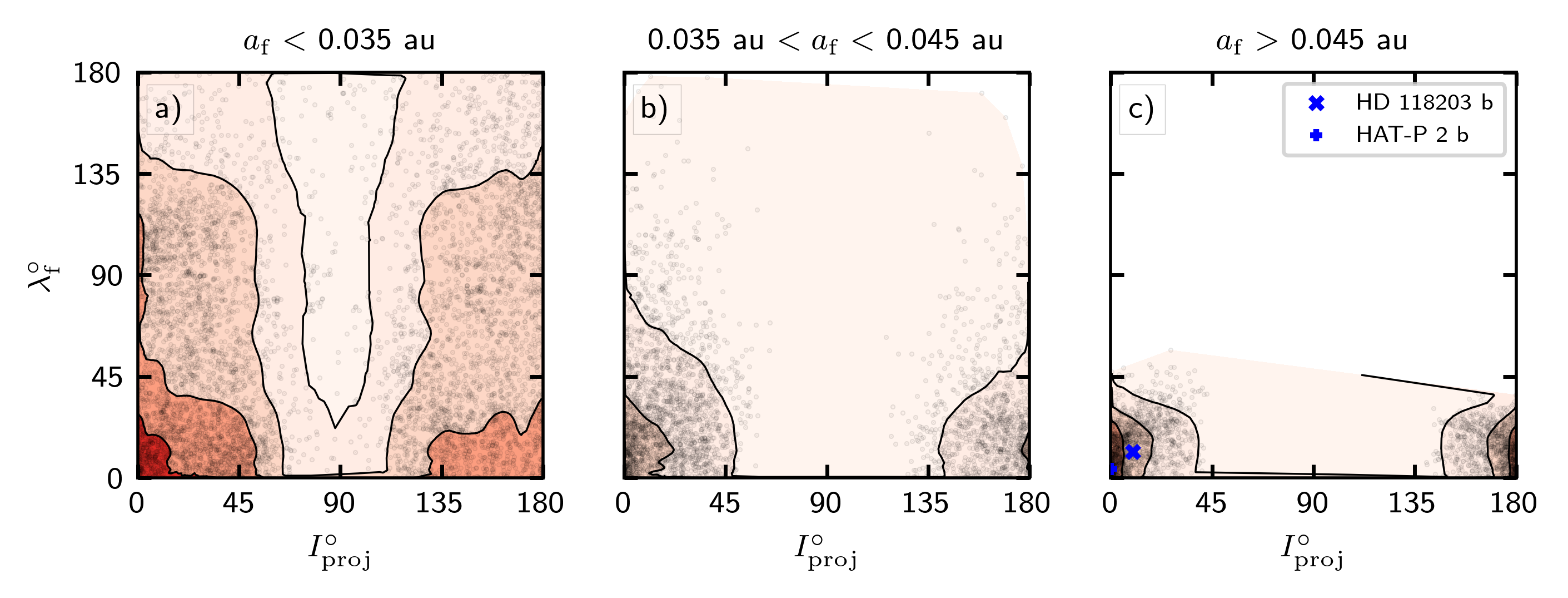}
	\caption{Projected obliquities and inclinations of HJs from our simulations. The x-axis shows the final projected mutual inclination between the two planets in the system, while the y-axis shows the projected stellar obliquity of the inner planet. Panel (a) shows results for HJs with final semi-major axes \(a_{\mathrm{f}} < 0.035\,\mathrm{AU}\). Panels (b) and (c) show results for \(0.035\,\mathrm{AU} < a_{\mathrm{f}} < 0.045\,\mathrm{AU}\) and \(a_{\mathrm{f}} > 0.045\,\mathrm{AU}\), respectively. The measured orbital properties of HD 118203 b and HAT-P-2 b are shown in panel (c). Each panel exhibits a distinct distribution in projected obliquity--inclination space.}
    \label{fig:pobqincaf}
\end{figure*}


\section{Discussion}
\label{sec:discuss}

The results above demonstrate that PPSHEM can reproduce the observed correlation between stellar obliquity and orbital separation without invoking post-migration tidal realignment. In our simulations, the final circularization radius is not merely a byproduct of tidal evolution but instead serves as a tracer of the secular pathway that drives the planet to high eccentricity. Systems undergoing rapid migration through quadrupole-order ZLK oscillations tend to reach smaller final orbital separations and exhibit a broad distribution of stellar obliquities. In contrast, systems driven by slower, octupole-order CHEM-like evolution typically circularize at larger orbital distances while remaining preferentially aligned. The resulting dependence of $\psi_{\rm f}$ on $a_{\rm f}$ emerges naturally in our fiducial simulations and remains robust under plausible reweightings of the initial eccentricity and inclination distributions.

We now discuss the implications of this interpretation for the observed HJ population, its predictions for mutual inclinations and migration timescales, its connection to other flavors of HEM, and the constraints that distant companions can provide.

\subsection{Best global fit to the data and scattering outcomes}

Our weighted ensembles show that the observed obliquity--period trend emerges only when the initial eccentricity and mutual-inclination distributions are sufficiently broad. If either $\sigma_e$ or $\sigma_I$ is too small, PPSHEM becomes inefficient and explores only a limited region of secular phase space. Moreover, the uniform grid adopted in our fiducial simulations is unlikely to represent a realistic population. Instead, planet--planet scattering naturally produces approximately Rayleigh-like eccentricity and inclination distributions \citep{juricTremaine2008,chatterjee}, motivating the weighting schemes explored in \S~\ref{sec:ICs}.

Among the models considered, the intermediate case shown in panel (e) of Figure~\ref{fig:afpsifweights} provides the best overall match to the data. It reproduces both the close-in population with a broad range of stellar obliquities and the wider-orbit population that remains preferentially aligned. At the same time, it yields a pile-up at periods of a few days and an outer-companion eccentricity distribution broadly consistent with observations \footnote{It should be noted that in the intermediate case, while the $a_{\mathrm{f}}-\psi_{\mathrm{f}}$ correlation remains robust, the HJ formation rate depends on the inclination distribution. For $\sigma_{\mathrm{e}}=0.3$, the formation rate increases from 0.03 for $\sigma_{\mathrm{I}}=25^\circ$ to 0.07 for $\sigma_{\mathrm{I}}=35^\circ$}. This preferred model does not correspond to either the pure ZLK or pure CHEM limits, but instead samples the continuum of secular pathways connecting them.

This result is consistent with the scattering experiments of \citet{Garzon2022}, who found that PPSHEM is fed through a diverse range of initial configurations. In their simulations, many systems enter the secular migration phase with moderate-to-large eccentricities and inclinations, rather than occupying exclusively the canonical ZLK or CHEM regimes.

The agreement is nevertheless not perfect. Even our best-fitting model underproduces short-period, low-obliquity HJs relative to the observed sample. One possible explanation is tidal dissipation within the host star, which becomes increasingly effective at the shortest orbital periods and may realign a fraction of these systems after migration \citep{Zanazzi2024}. Exploring the interplay between PPSHEM and post-migration tidal realignment is beyond the scope of this work and is left for future study.

\subsection{Correlation between projected mutual inclinations, obliquities, and semi-major axis}

A key prediction of PPSHEM is that the mutual inclination, stellar obliquity, and final orbital separation of HJs are correlated. In particular, systems undergoing CHEM retain low mutual inclinations throughout their evolution and produce HJs on relatively wide orbits. In contrast, systems driven by ZLK oscillations typically migrate to smaller orbital separations and exhibit broad distributions of both stellar obliquities and mutual inclinations.

This correlation is illustrated in Figure~\ref{fig:pobqincaf}. Similar to stellar obliquities, observations often constrain only the sky-projected mutual inclination between planetary orbits. We therefore present our results in terms of the projected mutual inclination, $I_{\rm proj}$, and the projected stellar obliquity, $\lambda$, of the HJs.\footnote{Projected inclinations and obliquities are calculated using Eq.~11 of \citet{FabryckyWinn2009}. We sample the stellar-spin azimuthal angle $\Omega$ uniformly between 0 and $2\pi$ and adopt an orbital inclination of $90^\circ$.}

Panel (a) shows the close-in HJ population. These systems exhibit broad distributions in both projected obliquity and projected mutual inclination, consistent with migration driven by ZLK oscillations. In contrast, the wider-orbit HJs shown in panel (c) are confined to a narrow region of parameter space, with both $\lambda$ and $I_{\rm proj}$ remaining relatively small. This behavior reflects their origin through CHEM, which excites large eccentricities while preserving near-coplanarity.

The intermediate population shown in panel (b) bridges these two regimes. The projected obliquities span a wide range of values, while the projected mutual inclinations remain comparatively constrained, typically below $\sim60^\circ$. Together, these results suggest that measurements of projected mutual inclinations in systems hosting outer companions may provide an additional diagnostic of the secular migration pathway responsible for producing HJs.

Hence, the mutual inclinations between HJs and their outer companions can help assess the relative importance of the ZLK and CHEM migration channels. At present, however, mutual inclinations remain poorly constrained in most observed systems. Nevertheless, \citet{beckerExteriorCompanionsHot2017} used statistical analyses of N-body integrations tailored to observed systems and found that the measured stellar obliquities are generally consistent with nearly coplanar planetary architectures.

A small number of systems already provide direct constraints on the orbital architecture through astrometric measurements. For example, HD~118203 hosts an eccentric HJ and an eccentric outer Jovian companion \citep{maciejewskiTrackingAdvancedPlanetary2024}. The inner planet has a line-of-sight inclination of $89^\circ$, while Gaia and Hipparcos astrometry constrain the outer companion to an inclination of $95^\circ$, implying a nearly coplanar configuration. Measurements of the Rossiter--McLaughlin effect combined with constraints on the stellar inclination further indicate a true stellar obliquity of $\sim30^\circ$ for the inner planet \citep{zhangTestbedTidalMigration2024}. Thus, HD~118203 occupies the region of parameter space predicted for CHEM-like migration, with both a low planet--planet mutual inclination and a modest stellar obliquity. We show its location in mutual-inclination--obliquity space in Figure~\ref{fig:pobqincaf}.

The HAT-P-2 system provides another possible example. This system hosts an eccentric HJ and a distant Jovian companion. The inner planet has a low projected stellar obliquity, while the astrometric non-detection of the outer companion places an upper limit on the mutual inclination between the planetary orbits \citep{debeursRevisitingOrbitalEvolution2023}. HAT-P-2 is likewise shown in Figure~\ref{fig:pobqincaf}.

In the near future, Gaia is expected to provide mutual-inclination constraints for many HJ systems with distant companions \citep[e.g.,][]{Espinoza-Retamal2023}. Combined with measurements of stellar obliquities, these observations will enable direct tests of the predicted correlation between HJ semi-major axis, stellar obliquity, and planet--planet mutual inclination, providing a powerful probe of the relative importance of different PPSHEM pathways.

Mutual inclinations may be especially valuable in systems where the present-day stellar obliquity is small. For many short-period HJs, tides may have partially erased the primordial spin--orbit misalignment by reorienting the stellar spin toward the planetary orbit \citep[e.g.,][]{Lai2012,Zanazzi2024}. In contrast, this process should leave the inclination between the HJ and its distant planetary companion largely unchanged. The planet--planet mutual inclination may therefore provide a more direct record of the migration history than the present-day stellar obliquity alone.

\subsection{Correlation between orbital period and migration timescale}

Our fiducial simulations predict a correlation between the final orbital period of a HJ and its migration timescale within the PPSHEM channel. Close-in HJs are preferentially produced by rapid, ZLK-driven migration, whereas wider-orbit HJs are more often associated with slower, CHEM-like evolution. Thus, within PPSHEM, the final orbital separation is not simply a consequence of tidal circularization, but also encodes the secular pathway and timescale through which the planet reached high eccentricity.

Recent observational studies provide an important population-level context for this prediction. Using stellar Galactic velocity dispersion as a relative age proxy, \citet{hamerEvidenceLateArrival2022} found evidence that misaligned HJs are older than aligned systems, suggesting that the mechanism responsible for producing large stellar obliquities operates at late times. More recently, \citet{schmidtMostHotJupiters2026} found that HJs inside and near the debiased peak of the period distribution are older than those outside the peak, and argued that a substantial fraction, $40$--$70\%$, of the HJ population must be produced by a late-time, peak-populating mechanism with a characteristic delay of $\gtrsim 1.5$~Gyr.

A direct comparison with our simulations requires some care. The observational age--period trend reflects the demographics of the entire HJ population, including possible contributions from in-situ formation, disk migration, planet--planet scattering, and different flavors of HEM. By contrast, our calculations isolate PPSHEM systems and focus on the secular delivery phase and subsequent circularization driven by tides raised on the planet. We do not follow the long-term post-migration tidal evolution driven by tides raised on the star, which can further shrink the orbits of the closest-in HJs and may contribute to the observed age ordering with $P_{\rm orb}$. In particular, stellar tides can move older near-peak systems into the inside-peak population over Gyr timescales, an effect not captured in our present integrations.

Therefore, the observational trend reported by \citet{schmidtMostHotJupiters2026} should not be interpreted as a one-to-one test of the PPSHEM migration timescales calculated here. Instead, it constrains the combined outcome of HJ formation, secular migration, and subsequent tidal evolution. In this broader picture, CHEM-like PPSHEM remains a viable contributor to the late-time, peak-populating component, especially because it naturally operates on long timescales and can populate systems near and outside the few-day period pile-up.

At the same time, the current observational sample remains too small, and the age constraints too population-averaged, to uniquely assign the late-time component to a specific secular pathway. A more direct test will require combining age information with additional diagnostics, such as stellar obliquities, outer-companion eccentricities, and planet--planet mutual inclinations. Such joint constraints will be needed to determine whether the late-time, peak-populating component inferred observationally is dominated by ZLK migration, CHEM, or a mixture of both.

\subsection{Comparison with other migration channels}

We now discuss the observed $a_{\rm f}$--$\psi_{\rm f}$ trend in the context of other proposed migration channels for HJs. In what follows, we focus primarily on high-eccentricity migration pathways. In its simplest form, disk-driven migration preserves spin--orbit alignment and therefore does not naturally explain the emergence of a strong correlation between stellar obliquity and final semi-major axis, unless additional mechanisms produce primordial star--disk misalignment.

\emph{Planet--planet scattering:} Planet--planet scattering can also generate HJs with a broad range of stellar obliquities \citep{nagasawaFormationHotPlanets2008,Nagasawa2011}. The simulations of \citet{beaugeMULTIPLEPLANETSCATTERINGORIGIN2012} show that HJs produced by rapid three-planet diffusion tend to end up on closer-in orbits and with broader obliquity distributions than systems that retain two planets. This trend has the same sign as the observed $a_{\rm f}$--$\psi_{\rm f}$ correlation. However, many planets in the three-planet scattering outcomes migrate too close to the star and are ultimately engulfed during subsequent tidal evolution.

The closest analogue to our scenario is provided by systems that relax to two planets after an ejection and subsequently undergo secular HEM. In these cases, the resulting HJs are predominantly prograde, although the closest-in planets show a modestly enhanced retrograde fraction. This suggests a possible correlation between final orbital separation and obliquity, but the trend appears weaker than in the secular pathways explored here. We note, however, that these integrations were evolved only up to 100 Myr and may therefore miss the late-arriving HJs expected from slower CHEM-like evolution.

\emph{Planet--planet secular HEM:}
Previous Monte Carlo experiments of PPSHEM by \citet{Petrovich2016} showed hints of an $a_{\rm f}$--$\psi_{\rm f}$ trend. However, those simulations sampled initial eccentricities only up to $e=0.3$ and adopted distributions with $\langle e\rangle \ll \langle I\rangle$. They therefore underrepresented systems that migrate through CHEM-like pathways, which require relatively large eccentricities and modest mutual inclinations. Thus, although these experiments can naturally generate the broad range of obliquities observed among HJs, the resulting PPSHEM population does not appear to produce as clear an $a_{\rm f}$--$\psi_{\rm f}$ correlation as found in our simulations.

\emph{Stellar ZLK mechanism:} Similar to PPSHEM, secular perturbations from a distant stellar companion can produce HJs with a wide range of stellar obliquities. The two mechanisms are closely related, but with an important distinction: there is no direct stellar analog of CHEM, since a massive, close-in, eccentric stellar companion would generally destabilize the proto-HJ orbit. As in this work, \citet{FabryckyTremaine2007} identified an envelope that limits the maximum final obliquity of a HJ at a given final semimajor axis. We derive an analytic expression for this envelope in \S~\ref{app:envder}. Including octupole-order perturbations increases the number of dynamical degrees of freedom, enabling orbital flips and broadening the resulting obliquity distribution \citep{LiCoplanarFlip2014}. This leads to a more diffuse distribution in the $a_{\rm f}$--$\psi_{\rm f}$ plane, as shown in Figure~\ref{fig:fidsimres}.

\emph{Secular chaos:} As in ZLK migration, planets undergoing secular chaos experience substantial inclination excitation and may therefore produce close-in HJs with a broad distribution of stellar obliquities, although the resulting orbits are expected to be predominantly prograde \citep{Lithwick2014,TeyssandierSecularChaos2019}. To our knowledge, existing population-synthesis studies of secular chaos have not reported a clear correlation between $\psi_{\rm f}$ and $a_{\rm f}$. We briefly explored whether adding an eccentric fourth planet to our simulations, thereby allowing for additional secular excitation and generally chaotic evolution, would erase the PPSHEM obliquity--semimajor-axis trend. The results are presented in Appendix~\ref{app:addplanets}. We find that the correlation between final semimajor axis and stellar obliquity identified in our two-planet simulations is largely preserved in the three-planet case. A more detailed investigation of higher-multiplicity systems is deferred to future work.

\subsection{Role of tidal dissipation}

In this work, we model tidal dissipation in both the star and the planet using the constant-time-lag tidal prescription. This simplified equilibrium-tide model assumes a weak tidal response and is characterized by a single parameter, the viscous timescale $t_{\rm v}$. To assess the sensitivity of our results to the assumed tidal efficiency, we repeated a subset of our fiducial simulations with a shorter planetary viscous timescale, $t_{\rm v,1}=0.1$~yr, corresponding to more efficient tidal dissipation. This experiment also provides a simple proxy for the possible role of dynamical tides, which are expected to become important at very large eccentricities, $e>0.9$, and can drain orbital energy more efficiently than equilibrium tides during close pericenter passages \citep{Vick2018,Wu2018,Vick2019}. These results are presented in Appendix~\ref{sec:tidemodel}.

As expected, reducing the planetary viscous timescale increases the hot Jupiter formation rate and decreases the tidal disruption rate. More efficient dissipation allows planets to circularize over a wider range of pericenter distances, thereby broadening the distribution of final semi-major axes. Importantly, however, the correlation between final semi-major axis and stellar obliquity persists, indicating that the $a_{\rm f}$--$\psi_{\rm f}$ trend is not an artifact of the particular tidal efficiency adopted in our fiducial model. Although a shorter equilibrium-tide viscous timescale cannot capture the detailed physics of dynamical tides, these experiments suggest that more efficient energy dissipation during high-eccentricity passages would likely strengthen HJ survival without erasing the main obliquity--separation trend.

\section{Conclusions}\label{sec:conc}

We have shown that planet--planet secular high-eccentricity migration naturally explains the observed correlation between hot-Jupiter orbital separation and stellar obliquity in single-star systems. 
In this picture, the final semi-major axis is a fossil record of both the secular pathway and the migration timescale. Highly inclined systems following von Zeipel–Lidov–Kozai-like evolutionary paths reach the HJ phase on comparatively shorter timescales, producing close-in HJs with broad obliquity distributions. More nearly coplanar systems evolve more slowly through coplanar high-eccentricity migration-like evolution, producing wider-orbit HJs that arrive later and remain preferentially aligned. These outcomes are not distinct populations, but limiting regimes of the same planet–planet secular migration process. Moderately broad initial eccentricity and inclination distributions populate the full continuum between these regimes, reproducing the observed $a_{\rm f}$–$\psi_{\rm f}$ trend while also broadly matching the HJ period distribution and the eccentricities of their outer companions.

This interpretation predicts that the three-dimensional architecture of HJ systems should vary with orbital separation. Close-in HJs should have broad mutual-inclination and obliquity distributions, while wider-orbit HJs should more often host nearly coplanar outer companions. Upcoming Gaia astrometric constraints on distant companions will provide a key test of this picture.

\section*{acknowledgments}
The authors are grateful to Hagai Perets, Songhu Wang, Gongjie Li, Billy Quarles, Doug Lin, Juliette Becker, Juan Ignacio Espinoza-Retamal, and Xian-Yu Wang for fruitful discussions. CP also thanks Dan Fabrycky for the initial encouragement to publish this work, which had been presented only schematically at the 2024 DDA meeting. This work was supported by the National Science Foundation under grants AST-2511257 and AST-2511258.

\appendix

\section{Set-up for Monte Carlo simulations}
\label{sec:setup_appendix}

We evolve two-planet systems orbiting a central star of mass \(m_0\). The planetary masses are denoted by \(m_1\) and \(m_2\), corresponding to the inner and outer planets, respectively. We use Jacobi coordinates to define the inner and outer orbits, whose orbital elements include the semi-major axes \(a_1\) and \(a_2\), eccentricities \(e_1\) and \(e_2\), and inclinations \(I_1\) and \(I_2\), respectively. For brevity, we often drop the subscript for the orbital elements of the inner planet.


To model the long-term evolution of the system, we adopt the secular approximation and focus on the hierarchical regime, characterized by a large separation between the planets (\( \alpha = a_1/a_2 < 0.1 \)). The dynamics are governed by a double-averaged Hamiltonian expanded to octupole order in \( \alpha \) \citep[e.g.,][]{fordSecularEvolutionHierarchical2000a, naozSecularDynamicsHierarchical2013}. Tidal dissipation is incorporated using the equilibrium tide model \citep[e.g.,][]{hutTidalEvolutionClose1981,eggletonEquilibriumTideModel1998a,eggletonOrbitalEvolutionBinary2001}, and we account for apsidal precession due to rotational and tidal deformation, as well as general relativistic effects. The equations of motion are adopted from \cite{petrovichSteadystatePlanetMigration2015} (Eqns A4–A9).

In our fiducial simulations, the masses of the host star and inner planet are fixed at \( 1\,M_\odot \) and \( 1 \,M_{\rm Jup} \) respectively, while the mass of the outer planet is sampled uniformly between 1 and 3 \( M_{\rm Jup} \). The radii of the host star and inner planet are fixed at \( R_1 = 1\,R_\odot \) and \( R_2 = 1\,R_{\rm Jup} \), respectively. We set the semi-major axes of the inner and outer planets to \( a_1 = 1 \) AU and \( a_2 = 10 \) AU, respectively. The arguments of pericenter and longitudes of ascending node for both planets are sampled uniformly between \( 0^\circ \) to \( 360^\circ \). The viscous dissipation timescales are set to \( t_{ \rm v,0} = 50 \) years for the star and \( t_{\rm v,1} = 0.1 \) years for the planet. The tidal Love numbers are taken to be \( k_{\rm 2,0} = 0.028 \) for the star and \( k_{\rm 2,1} = 0.5 \) for the inner planet. 

Given the sensitivity of the system's evolution to the initial eccentricities and mutual inclination, we sample these parameters extensively. The initial eccentricities are chosen from a uniform grid spanning \(0 \leq e_1, e_2 < 1\), with spacing \(\Delta e = 0.05\). The initial mutual inclination is sampled uniformly in \(I_{\rm mut}\), with a spacing of \(\Delta I_{\rm mut} = 3^\circ\). Before running each simulation, we verify the long-term stability of the system using the empirical criterion of \cite{Petrovich2015}.

\section{Derivation of obliquity envelope under test-particle quadupole order approximation}
\label{app:envder}

The dynamics of a planet undergoing secular high-eccentricity migration (HEM) are governed by two main effects: secular perturbations from the outer companion and tidal dissipation within the planet. At quadrupole order, the secular evolution conserves \(J_z = \sqrt{a(1-e^2)}\cos I_{\mathrm{mut}}\). Tidal dissipation within the planet, under the assumption of pseudosynchronous rotation, approximately conserves the orbital angular momentum \(J = \sqrt{a(1-e^2)}\). Moreover, pseudosynchronous tidal dissipation does not change the planet's inclination. Hence, \(J_z\) is approximately conserved under pseudosynchronous ZLK driven HEM.

During HEM, the proto-hot Jupiter exchanges angular momentum with an outer companion, exciting its eccentricity while leaving its semi-major axis approximately constant. During high-eccentricity passages, tidal dissipation approximately conserves the planet's orbital angular momentum while shrinking and circularizing the orbit. The planet eventually reaches a final semi-major axis \(a_{\mathrm{f}}\), at which point its eccentricity has been damped to nearly zero. Assuming conservation of orbital angular momentum during tidal circularization, the final semi-major axis is approximately \(a_{\mathrm{f}} \simeq a_{\rm init}(1-e_{\mathrm{max}}^2)\), where \(a_{\rm init }\) is the initial semi-major axis of the inner planet and \(e_{\mathrm{max}}\) is the maximum eccentricity attained during the ZLK cycle. Apsidal precession due to general relativity, tides, and rotational deformation can significantly affect \(e_{\mathrm{max}}\); we account for these effects using the analytical expression from \cite{liu15}.

As the planet migrates inward, it becomes more tightly coupled to the host star and gradually decouples from the outer companion. Consequently, the mutual inclination between the planetary orbits becomes effectively frozen. Using conservation of \(J_z\), the final mutual inclination is given by
\begin{equation}
I_{\mathrm{mut,f}}
=
\cos^{-1}
\left[
\sqrt{\frac{a_{\rm init}(1-e_{\rm init}^2)\cos I_{\rm mut,init}}{a_{\mathrm{f}}}}
\right].
\label{eq:final-mutual-inclination}
\end{equation}
Here \(e_{\rm init}\) and \(I_{\rm mut, init}\) are the initial eccentricity and mutual inclination between the planets, respectively. If the initial stellar obliquity is zero, the final stellar obliquity lies in the range \(I_{\rm mut, init} - I_{\mathrm{mut,f}} < \psi < I_{\rm mut,init} + I_{\mathrm{mut,f}}\). This condition constrains the maximum final obliquity of a HJ for a given final semi-major axis.

\begin{figure}
	\centering
	\includegraphics[scale=0.6]{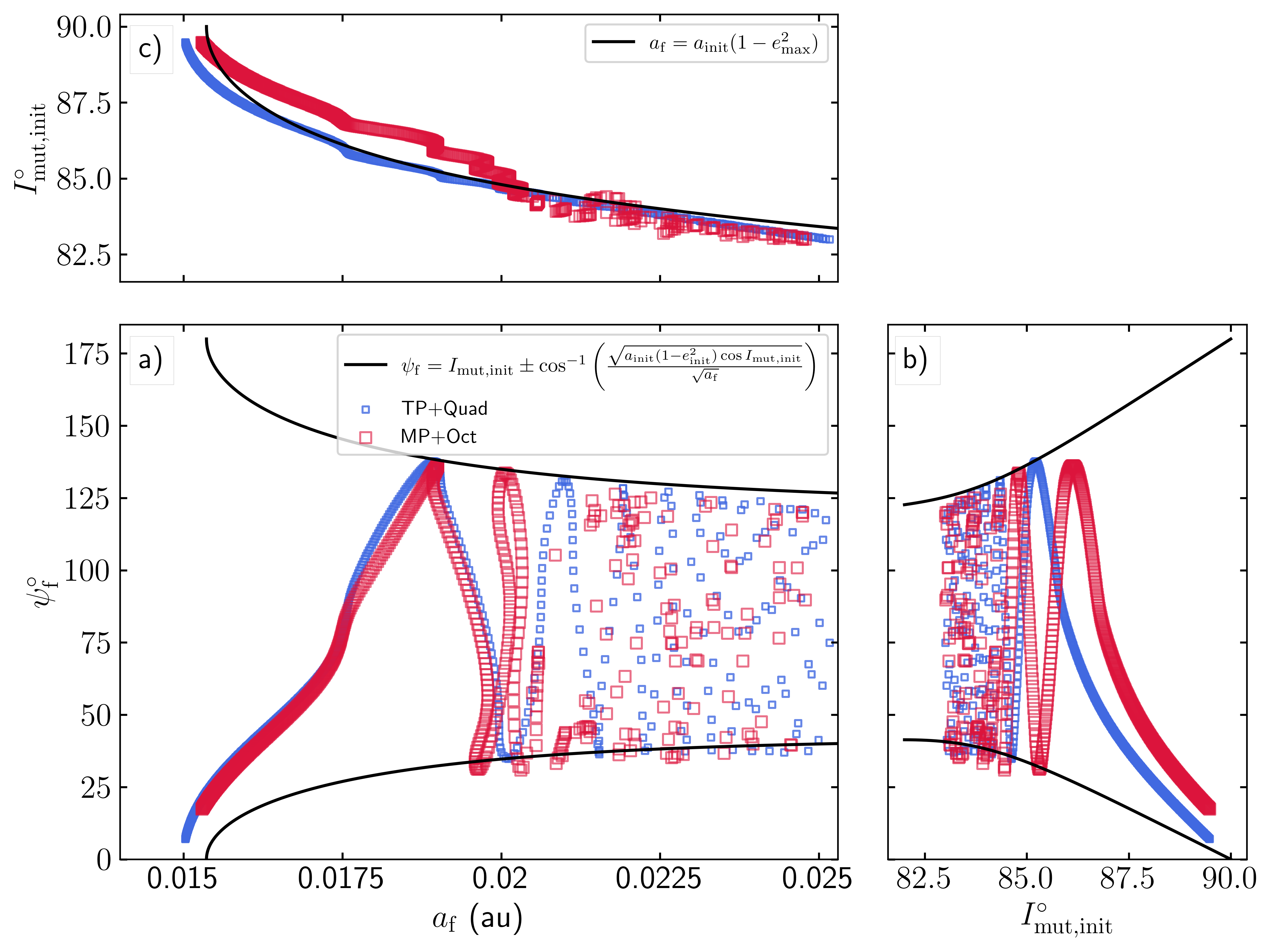}
	\caption{Comparison of analytical expression for obliquity envelope with secular simulations. In panel (a), we show the final semi-major axis on the x-axis, and the final obliquity on the y-axis. Panel (b) shows final obliquity as a function of the initial mutual inclination. In the panel (c), we show the initial mutual inclination between the planets as a function of the final semi-major axis of the HJ.  The analytical results are shown as solid black lines. The results from our secular simulations are shown as red and blue squares. The results shown in blue only take into quadrupole order perturbations, and assume that the inner planet is a test particle. For the results shown in red, test particle approximation is dropped, and octupole order perturbations are also included. We can see that the analytical envelope is in good agreement with both sets of secular simulations.}
	\label{fig:obqenv}
\end{figure}

A comparison of the above expression with secular simulations is shown in Figure \ref{fig:obqenv}. We can see that it is excellent agreement with the simulations. Even when the test-particle quadrupole order approximation is dropped, the above expression is qualitatively in agreement with the simulations. In summary, an envelope emerges because both the final semi-major axis and the the final obliquity of the HJ depend on the initial mutual inclination between the planets. As the mutual inclination increases, the inner planet can reach higher eccentricities, enabling the proto HJ to migrate to shorter orbits. A larger initial mutual inclination also drives inclination oscillations with larger amplitudes, allowing the obliquity to vary over a wider range.
 
\section{Dependence on the Tidal Dissipation Models}
\label{sec:tidemodel}
\begin{figure}
\centering
	\includegraphics[scale=0.9]{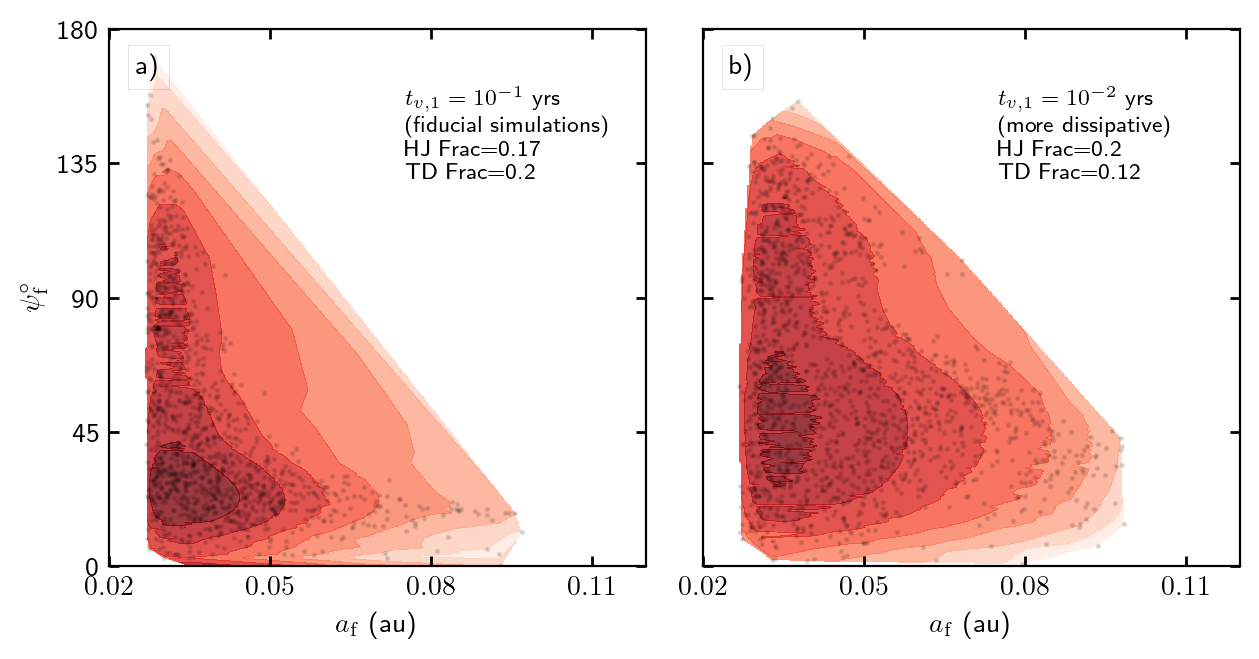}
	\caption{The final obliquity and semi-major axis of HJs in our secular simulations. The black dots show results from our simulations, and the contours show the two dimensional kernel density estimate. Results from our fiducial simulations (panel a) are compared to an additional set of simulations in which the tidal dissipation in the planet is enhanced (panel b). Rest of the initial conditions are kept same. We can see that the results are qualitatively similar, except stronger tides tend to produce more HJs on wider orbits.}
    \label{fig:afpsiftd}
\end{figure}
The tidal dissipation rate within the planet regulates both the migration timescale and the final semi-major axis of a hot Jupiter. Observations of giant planets in the Solar System suggest tidal quality factors of $Q \sim 10^{5-6}$ \citep{Goldreich1966}. For gaseous exoplanets, the tidal quality factor has been estimated to be in the range $10^6 < Q < 10^8$ \citep{Jackson2008,Hansen2010,Hansen2012,Quinn2014,Fellay2023}. The viscous dissipation timescale $t_{v,1} = 0.1\,\mathrm{yr}$ used in our fiducial simulations is broadly consistent with these constraints, although significant uncertainties remain in the tidal parameters of HJs.

We ran additional simulations in which we enhance the tidal dissipation in the inner planet. We show the effects of stronger tidal dissipation in panel (b) of Figure \ref{fig:afpsiftd}. We can see that the results are qualitatively similar to our fiducial simulations; we recover the envelope which constrains the range of obliquities HJs can have at a given semi-major axis. This is expected since the correlation between obliquities and semi-major axes of HJs is primarily driven by the secular three body dynamics. Stronger tides primarily tend to produce HJs with a broader range of final semi-major axes.

It should be noted that observed sample shown in Figure \ref{fig:ensa1fpsif} has a longer tail of low obliquity wide orbit HJs as compared to our fiducial simulations shown in Figure \ref{fig:fidsimres}. From the above analysis, we can see that the distribution of final semi-major axis of HJs depends on the efficiency of tidal dissipation in the planet. Overall, our model is consistent with observations within the uncertainities in the tidal models.

\section{Additional Planets}
\label{app:addplanets}

\begin{figure*}
    \centering
    \includegraphics[width=1.0\linewidth]{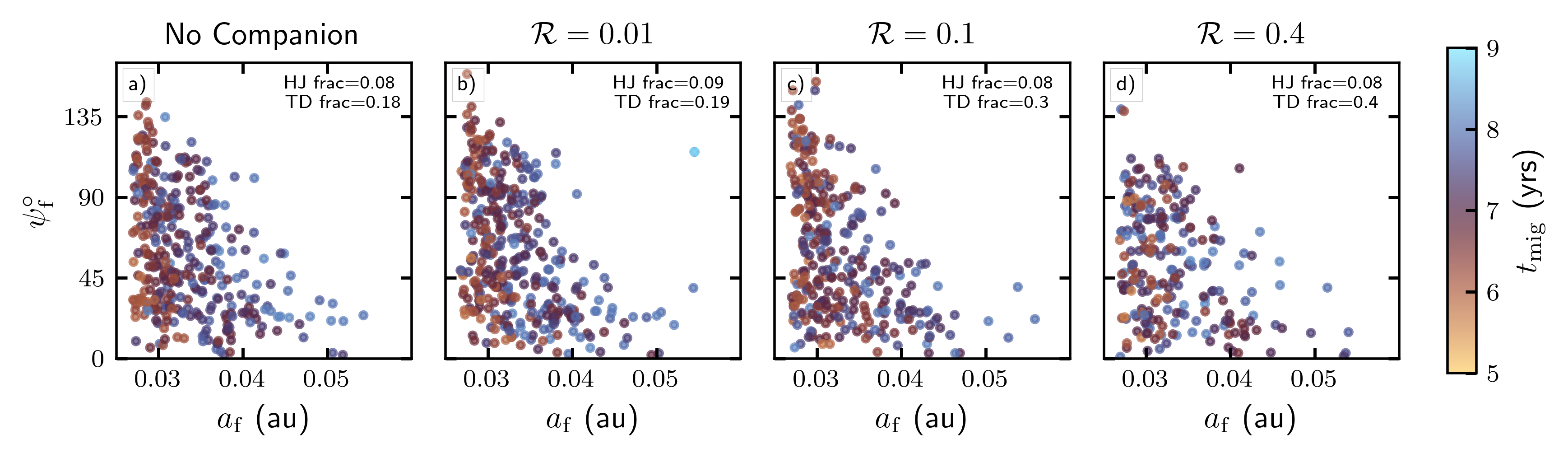}
    \caption{Distribution of final obliquities and semi-major axes of HJs in our three-planet simulations. Panel (a) corresponds to our fiducial setup. In panels (b), (c), and (d), we include an additional outer planet with mass \(8\,M_{\mathrm{Jup}}\). The location of the outermost planet is set by the parameter \(\mathcal{R}\) (see text for details): it lies farthest out for \(\mathcal{R}=0.01\) and closest in for \(\mathcal{R}=0.4\). Panels (a) and (b) are similar, indicating that a distant outer planet does not significantly alter the final distribution of semi-major axes and obliquities. For \(\mathcal{R}=0.1\) and \(\mathcal{R}=0.4\), all planets in the system are tightly coupled. Even in these cases, we recover the envelope relating obliquity to semi-major axis, demonstrating the robustness of our fiducial results.}
    \label{fig:fourplanetafpsif}
\end{figure*}
Planet-planet scattering experiments show that most unstable multi-planetary systems relax into two planet configurations. However, in a non-negligible fraction of cases, a third planet can remain bound to the system after early instability phase \citep[e.g.,][]{beaugeMULTIPLEPLANETSCATTERINGORIGIN2012}. When three planets survive, they settle into a hierarchical configuration that allows long‑term stability and secular evolution. The dynamics of such a system is dictated by the parameter $\mathcal{R}\equiv \frac{m_3}{m_2}\left(\frac{a_2}{a_1}\right)^{\frac{3}{2}}\left(\frac{a_2}{a_3}\right)^{3}\frac{(1-e_2^2)^{\frac{3}{2}}}{(1-e_3^2)^{\frac{3}{2}}}$ \citep{Hamers2016}. If the outer most planet resides on a very wide orbit ($\mathcal{R}<<1$), it is unlikely to affect the dynamics of the inner two planet system, and the resulting dynamics is similar to our fiducial simulations. When $\mathcal{R}\sim1$, the long term evolution of all three planets becomes coupled and is generally chaotic.

We ran additional set of secular four body simulations in which we placed a third planet at different distances from the outer planet of our fiducial simulations. We use equations of motions for 3+1 hierarchical systems derived in \citet{Hamers2015}, and because we are working in the planetary regime, ignore the cross terms. The semi-major axis of the outermost planet was chosen such that $\mathcal{R}=\{0.4,0.1,0.01\}$. It's eccentricity was fixed to 0.6, and it's orbit was initialized on the same plane as the middle planet. To identify stable configurations, we applied the empirical stability criteria of \citet{Petrovich2015} on each pair of adjoining planets.

Around $20\%$ of the systems from our fiducial setup were deemed unstable in a $\mathcal{R}=0.4$ configuration. Interestingly, while the number of planets which do not migrate decreases with $\mathcal{R}$, HJ formation rate does not change. This difference arises from a substantial increase in tidal disruptions. Around $40\%$ of the systems experience tidal disruptions in our $\mathcal{R}=0.4$ simulations, compared to only around $18\%$ in our fiducial simulations.

Figure~\ref{fig:fourplanetafpsif} shows the results of our secular simulations with three planets, except for panel~(a), which uses a setup identical to our fiducial two-planet simulations. The simulations with $\mathcal{R} = 0.01$ are effectively indistinguishable from the fiducial case, as expected, since the perturber remains far from the inner two planets in this configuration.

For $\mathcal{R} = 0.1$ and $\mathcal{R} = 0.4$, we still recover the characteristic envelope that constrains the obliquity as a function of the semi-major axis, even though the outer companion is more tightly coupled to the inner planets. In particular, in the $\mathcal{R} = 0.4$ system, a substantial fraction of planets that would otherwise have migrated to become HJs instead experienced tidal disruption (about $60\%$ of cases) or were classified as dynamically unstable (about $16\%$). Tidal disruptions occurred primarily for planets undergoing ZLK-driven high-eccentricity migration.

In summary, hierarchical three-planet systems exhibit markedly more chaotic long-term evolution, accompanied by a substantially increased rate of tidal disruption. Nonetheless, the planets that do become HJs display $a_\mathrm{f}$–$\psi_\mathrm{f}$ correlation similar to our fiducial two-planet simulations.

\bibliography{refs,obqpaper,ganHJobq,diego_refs,CP_refs}

\end{document}